\documentclass[preprint,showpacs,preprintnumbers,amsmath,amssymb,eqsecnum,aps]
{revtex4}

\usepackage{epsfig}
\usepackage{graphicx}
\usepackage{dcolumn}
\usepackage{bm}

\begin{document}

      \title{Acceleration of the universe, vacuum metamorphosis,
                       and the large-time
		asymptotic form of the heat kernel
}


\author{Leonard Parker\footnote{E-mail address: leonard@uwm.edu}
 and
Daniel A.\ T.\ Vanzella\footnote{Present 
address: Instituto de F\'\i sica, 
Universidade de S\~ao Paulo, C.P.~66318,
S\~ao Paulo, SP, 05315-970, Brazil. 
E-mail address: vanzella@fma.if.usp.br} }
\affiliation{Physics Department, University of Wisconsin-Milwaukee,\\
        P.O.Box 413, Milwaukee, WI 53201}


\begin{abstract}
We investigate the possibility that the late 
acceleration observed in the rate of expansion of the universe
is due to vacuum quantum effects arising
in curved spacetime. The theoretical
basis of the vacuum cold dark matter (VCDM),
or {\it vacuum metamorphosis}, cosmological 
model of Parker and Raval is revisited and improved.
We show, by means of a manifestly nonperturbative 
approach, how the infrared behavior of the 
propagator (related to the large-time asymptotic form 
of the heat kernel) of a free scalar field in 
curved spacetime leads to nonperturbative terms in the 
effective action similar to those appearing in the 
earlier version of the VCDM model. The asymptotic form that we 
adopt for the propagator or heat kernel at {\it large} proper 
time $s$ is motivated by, and consistent with, particular cases
where the heat kernel has been calculated {\it exactly}, 
namely in de Sitter spacetime, in the Einstein static
universe, and in the linearly-expanding 
spatially-flat FRW universe. This large-$s$ asymptotic
form generalizes somewhat the one suggested by
the Gaussian approximation and the $R$-summed form of 
the propagator that earlier served as a theoretical
basis for the VCDM model.
The vacuum expectation value for the energy-momentum tensor
of the free scalar field, obtained through variation of 
the effective action, exhibits a resonance effect when the 
scalar curvature $R$ of the spacetime reaches a 
particular value related to the mass of the field.
Modeling our universe by an FRW spacetime filled with 
classical matter and radiation, we show that 
the back reaction caused by this resonance
drives the universe through a transition to an accelerating 
expansion phase, very much in the same way as originally 
proposed by Parker and Raval. Our analysis includes higher
derivatives that were neglected in the earlier analysis, and
takes into account the possible runaway solutions that can
follow from these higher-derivative terms. We find that the
runaway solutions do not occur if the universe was described by
the usual classical FRW solution prior to the growth of vacuum 
energy-density and negative pressure (i.e., vacuum metamorphosis)
that causes the transition to an accelerating expansion of the 
universe in this theory.

\end{abstract}
\maketitle

\section{Introduction}
\label{intro}

Recent observations of type Ia supernovae (SNe-Ia) appear to support the 
view that the expansion of the universe started speeding 
up about 5 Gyr before the present time, and that it was 
slowing down prior to that time~\cite{Perl1}-\cite{Riess2}. 
Together with the evidence from the power 
spectrum of the cosmic microwave background 
radiation (CMBR)~\cite{Bernardis}-\cite{Spergel}
establishing that the universe
is nearly spatially flat, these observations imply that
a significant fraction of the energy density of the universe
is in the form of a dark energy that exerts negative pressure.
The focus of this paper will be on one possible candidate for
this dark energy.

It has been suggested~\cite{PR1, PR2345} that the 
recent acceleration of the universe may be caused by the vacuum
energy and pressure of a quantized scalar field in
curved spacetime. This scalar field is assumed to be free,
meaning that it has no non-gravitational interactions.
Furthermore, the mass $m$ of the field is assumed to have a very
small non-zero value, so the Compton wavelength of the particle 
is a significant fraction of the present Hubble radius of the
universe. The origin of the acceleration in this theory is
not the zero-point energy that would give rise to a cosmological
constant $\Lambda$. The latter would be affected by renormalization
and would be expected to be very large, if it is not forced to
be zero or nearly zero by some symmetry, dynamical process, or
other principle that is not well understood. The curved spacetime 
effect considered here has a magnitude that depends on the 
ratio of the mass of the free scalar field to
a dimensionless constant of order unity. For a free scalar field
in curved spacetime, the mass of the field is not 
renormalized~\cite{BP,PT2}.
Hence, the mass of this free field may be taken to be very small 
and will remain small within the context of quantum field theory 
in curved spacetime. Thus, the theory studied here, with a 
very small value for $m$, is internally consistent.

The gravitational field is treated as a classical field here.
One may ask if quantizing the gravitational field,
by regarding gravitons as quantized fluctuations of the metric
on the curved spacetime background, would necessarily force 
the effective mass $m$ of the scalar field to be large?
The self-energy contribution to the scalar field propagator
caused by virtual gravitons is evidently not renormalizable,
so a cut-off must be introduced. The magnitude of the
effective mass of the scalar particle would then depend on
the cut-off. If the cut-off is somewhat smaller than the
Planck mass, then the graviton contribution to the effective
mass would appear to be small. This would not affect
our cosmological predictions because they depend on a
quantity $\bar{m}^2$, of order $m^2$, that is determined
by the cosmological data. We make no attempt here to
obtain the value of $m$ from first principles.

The VCDM model gives a satisfactory fit to the observed
CMBR power spectrum and SNe-Ia data~\cite{PKV}. It may also
lead to the observed suppression of the CMBR power spectrum
at very low values of {\it l}~\cite{CPVprep}.

The question of how the very small mass scale fits in with 
fundamental theories of elementary particles and strings is 
not one that has a clear answer. It has been 
suggested~\cite{FHSW} that a scalar particle of
the required small mass may arise via symmetry breaking
as a pseudo-Nambu-Goldstone boson (pngb). It seems possible that
such a pngb, when quantized in curved spacetime, could lead to 
similar effects in curved spacetime through an effective 
action~\cite{CaldwellParker} or other methods~\cite{SahniHabib}.
In the present paper, only the free field will be considered.

In the previous work of Parker and 
Raval~\cite{PR1,PR2345}, 
the approximation was made
that the propagator (the analytic continuation of the heat kernel)
was proportional to 
$\exp\left[-i \left(m^2 + \left(\xi - 1/6 \right) R\right)s \right] 
\left( 1 + (i s)^2 \bar{f}_2 \right)$, where $m$ is the mass of the
free scalar field, $R$ is the scalar curvature, $\xi$ is 
a dimensionless coupling constant that appears
in the equation governing the free scalar field in curved spacetime,
and $\bar{f}_2$ is a quantity constructed from 
covariant derivatives of $R$ and contractions of
products of two Riemann tensors [see Eq.~(\ref{f2bar})]. 

This approximation was obtained
from the expansion of the heat kernel in powers of the proper-time
parameter $s$, with the exponential factor involving $R$
coming from summing all terms that involve one or more factors
of $R$ to all orders in 
$s$~\cite{PT,JP}.
Alternatively, the exponential involving $R$ can be
obtained by means of a Feynman path integral solution of the
Schr{\" o}dinger equation for the propagator~\cite{BeP}.

The presence of the exponential in $R$ was shown by Parker and Raval
to lead to a growth in $\langle T_{\mu\nu}\rangle$, the 
expectation value of the energy-momentum tensor of the scalar
field, when $R$ falls to a value of the order of $m^2$. The reaction
back of this growth of $\langle T_{\mu\nu}\rangle$ in the
semi-classical Einstein equations was shown to cause the expansion
of the universe to accelerate in such a way as to keep the
scalar curvature, $R$, of the spacetime nearly constant.

There are several natural questions that one can raise about
the previous approach. For example,
one may question the validity of the above approximation
to the heat kernel, and in particular the use of that
form of the heat kernel for large values of $s$. It is the
large-$s$ form of the propagator that gives rise to the
terms in $\langle T_{\mu\nu}\rangle$ that grow large as $R$ 
approaches a critical value of order $m^2$. A partial
justification is that the factor 
$\exp\left[-i \left(m^2 + \left(\xi - 1/6 \right) R\right)s \right]$
in the propagator comes from all powers of $s$, and is 
thus nonperturbative. However, the
series in powers of $s$ is in general asymptotic, not convergent,
so the convergent subset of terms involving $R$ that are
summed may not fully reflect the form of the propagator at
large $s$. It should be mentioned
that the Gaussian approximation to the path integral of
Bekenstein and Parker~\cite{BeP} is independent of the power 
series in $s$ and gives reason to expect an exponential factor at
large $s$. However, it is also an approximation, and thus leaves 
room for some modification. 

In the present paper, we directly examine
the large-$s$ asymptotic form of the heat kernel for several
cases in which the heat kernel is {\it exactly} known.  We postulate
a general expression for the asymptotic form that is consistent 
with the cases studied. We show that the transition to
constant $R$ occurs as a result of the large-$s$ asymptotic
form of the heat kernel. In our numerical integration
of the Einstein equations in the FRW universe, we use the
postulated asymptotic form with a particular choice of a factor that 
is quadratic in the Riemann tensor. This choice (namely, $\bar{f}_2$)
is consistent with the exact asymptotic forms considered, 
but there are also other consistent possibilities, 
as we discuss. We expect that the transition is fairly 
generic, as the choice we make of the quadratic factor 
is sufficiently complicated to be representative. As noted
below, the possible runaway solutions of the field
equations obtained from this higher derivative effective
action do not occur with physically acceptable initial conditions.
This approach based on the asymptotic form of the heat kernel is
general enough that it may be useful in other fundamental 
theories, such as string theory, that have higher derivative 
terms in their low-energy effective actions.

One may also question the validity, in the earlier work of 
Parker and Raval, of neglecting
derivatives of the Riemann tensor in
arriving at the expression for $\langle T_{\mu\nu}\rangle$.
The reason for neglecting such derivatives 
was that the present expansion of the universe 
is slow enough that terms involving derivatives 
of invariants were expected to be small with respect to 
other terms in the effective action that do not involve such 
derivatives. However, the solution for the expansion of the universe
that they found went through a relatively rapid transition from
a matter-dominated expansion to a constant-$R$ expansion. During
that transition it may be necessary to include the derivative terms
that were neglected. Furthermore, by including the higher
derivative terms one may generate unstable runaway solutions
of the Einstein equations. 

In the present work, we keep the 
derivative terms that were neglected in the earlier work. Then
the covariant conservation of $\langle T_{\mu\nu}\rangle$ is
satisfied at all times without neglecting derivatives. We find
that the transition to constant $R$ still occurs. We also
carefully consider runaway solutions. We find that if the
early evolution, prior to the transition, is described by
the usual classical solution to the Einstein equations 
(as one expects because $\langle T_{\mu\nu}\rangle$ is 
negligible before the transition), then there is a transition
to the constant $R$ solution with no runaway solutions. There are
runaway solutions only if one takes the initial evolution to be
significantly different from that of the classical solution 
prior to the transition.

We also generalize the previous work by including
the neglected effect of dissipative processes. These could come
from very small interactions of the scalar field $\phi$ with
other fields, which could involve dissipative processes such
as very slow decay of the scalar particles into less massive fields, 
or weak radiative processes, such as the
production of gravitational waves. In the present paper, we
model such dissipation phenomenologically by introducing a  small,
but nonzero, imaginary part in the mass term.

In their previous work, Parker and Raval mentioned the
possibility that the mechanism they considered may be relevant
to early inflation. However, because the series in powers of $s$
is more difficult to justify as an approximation when the curvature
terms are large, they applied their work only to the recent universe.
However, now that the mechanism has been further justified by 
considering known {\it exact} asymptotic forms of the heat kernel, 
the time may be ripe to apply this mechanism to {\it large-mass} 
scalar fields that are present in elementary particle theories. 
The acceleration effect will still come from the large-$s$ 
behavior of the heat kernel using the same postulated
asymptotic form. The fact that the mass is large would not seem 
to prevent the growth of $\langle T_{\mu\nu}\rangle$. Because 
of the large mass, these bosons could give rise to 
early inflation. At the very least, our mechanism could 
modify the considerations of early inflation that 
come from possible self-interaction potentials of 
these fields. It is also possible that our mechanism
could give rise to a satisfactory new model of early inflation 
in the absence of self-interaction potentials. Thus, this type of
field might serve as a candidate for an inflaton. The dissipative 
term would then be larger because of the interactions and decay 
channels of the massive scalar boson. One would
of course have to consider the questions of reheating and the
perturbation spectrum produced in such an inflationary epoch.
This mechanism for early inflation is worthy of investigation,
but it is not our main focus here, and will not be considered 
further in the present paper.

The very low-mass, free, scalar particle that we consider 
in this paper as a possible cause of the {\it recent} acceleration 
of the universe, may appropriately be called an {\it acceletron}, 
to distinguish it from an inflaton. Even in the absence of other
mechanisms for early inflation, the acceletron itself would
produce an early inflation at the Planck scale, similar 
to Starobinsky inflation~\cite{Starobinsky}. Assuming that
a successful exit from early inflation occurs, then at a much
later time, the same acceletron would produce the presently
observed acceleration of the universe.

In Section~\ref{sec:formalism}, we outline 
the basic theoretical structure, including
the regularized expression for the effective action.
In Section~\ref{sec:partcases}, we discuss the  asymptotic form of the
heat kernel for large-$s$, and make a postulate for this
asymptotic form that is consistent with the asymptotic
forms of several exact solutions for the heat kernel.
In Section~\ref{sec:general}, we show 
that the large-$s$ asymptotic form of the
heat kernel is what gives rise to the terms 
that make large contributions to the vacuum energy-density and
corresponding negative pressure when the scalar curvature falls
to a magnitude comparable to the square of the mass of the
acceletron field, $\phi$.
In Section~\ref{sec:renormvar}, 
we write the renormalized effective action, and
give the corresponding energy-momentum tensor.
The relevant variations, including terms with derivatives of
the Riemann tensor, are given in Appendix~\ref{appa}.
In Section~\ref{sec:vcdm}, we briefly summarize results of numerical
integration of the Einstein equations in an FRW cosmological
spacetime. More detailed results of the numerical integration 
will be given in a later paper that is now in preparation.
In Section~\ref{sec:discussion}, we present our conclusions.


\section{Effective-action formalism}
\label{sec:formalism}

In this section we outline the method
of relating the one-loop effective action of a
quantum field in curved spacetime to the 
generalized $\zeta$-function 
and the propagator in the proper-time formalism.

The action $S$ of an uncharged scalar field $\phi$ in curved
spacetime can be written in the form
\begin{eqnarray}
S=S[g_{\mu \nu},\phi]\equiv
-\frac{1}{2}\int d^4x\sqrt{-g} \,\phi(x)H(x)\phi(x),
\label{ca}
\end{eqnarray}
with $H(x)$ being the differential operator
\begin{eqnarray}
H(x)\equiv -\square +m^2+\xi R,
\label{H}
\end{eqnarray}
$\square \equiv g^{\mu \nu}\nabla_\mu \nabla_\nu$,
and $\xi$ being a coupling constant between the field $\phi$ and the 
scalar 
curvature $R=R(x)$ of the spacetime.

The corrections to the classical action of the gravitational
field that are caused by quantum fluctuations of $\phi$ can be
found by evaluating the effective action $W_q=W_q[g_{\mu \nu}]$.
The latter is defined for a given spacetime with 
metric $g_{\mu \nu}$ through the 
functional integral
\begin{eqnarray}
e^{iW_q}=\int d[\phi]\,e^{iS}.
\label{W1}
\end{eqnarray}
In analogy with the Feynman path integral, the functional
integral is proportional to a probability amplitude to go
from an ``in'' state to an ``out'' state determined by
the configurations of $\phi$ that one sums over. Rather
than explicitly specifying the initial and final configurations
of $\phi$, the ``in'' and ``out'' states are specified implicitly
by boundary conditions used later in evaluating the 
functional integral.
[To insure convergence of the functional integral, the operator
$H(x)$ is understood to be $H(x) - i\epsilon$, where $\epsilon$ is a 
positive infinitesimal.]

A nonzero imaginary part of $W_q$ implies
a nonzero rate of particle production of this
scalar field~\cite{Schwinger}.
If the spacetime is such that we can neglect the imaginary
part of $W_q$, then the ``in'' and ``out'' states
of the $\phi$ field are equivalent, differing only by a phase 
factor. The expectation value of the energy-momentum tensor of the 
scalar field in this state is then given by
\begin{equation}
\langle T^{\mu\nu} \rangle = 
         \frac{2}{\sqrt{-g}} \frac{\delta W_q}{\delta g_{\mu\nu}}.
\label{expectTmn}
\end{equation}
In the semi-classical theory, this expectation value appears on 
the right-hand-side of the Einstein gravitational field equations, 
along with the other sources of the gravitational field. Later, we
will apply this formulation to a spacetime for which it appears quite
suitable; namely, the Friedmann-Robertson-Walker (FRW) universe 
describing the averaged behavior of the {\em recent} expansion 
of the universe.

Because the action $S$ of the free scalar field is quadratic in $\phi$,
the functional integral reduces to a Gaussian integral which gives
(to within a constant normalization factor that does not affect the
variation of $W_q$):
\begin{equation}
e^{i W_q} = 
  1/\sqrt{{\rm Det}\left({\hat H} / \mu^2 \right)}\, ,
\end{equation}
where $\hat H$ is the abstract operator defined through
\begin{eqnarray}
\left<\left.x\right|\right.{\hat H}\left.\left|x'\right.\right>
=
H(x)\delta(x,x')\;,
\label{HhatHx}
\end{eqnarray}
with $\left.\left| x \right.\right>$ being position eigenstates
normalized through
\begin{eqnarray}
\left< x \vert x'\right>
=\delta(x,x')\equiv \frac{\delta^4(x-x')}{\sqrt{-g}}\;.
\label{ketx}
\end{eqnarray}
Then
\begin{equation}
W_q = (i/2) \ln {\rm Det}\left({\hat H}/\mu^2 \right) =
    (i/2) {\rm Tr} \ln \left({\hat H}/\mu^2 \right),
\label{W1Hhat}
\end{equation}
with $\mu$ an arbitrary constant having the units of mass.
The renormalized effective action will not depend on $\mu$.

The expression for $W_q$ can be regularized by writing it
in terms of the generalized $\zeta$-function~\cite{DC0,Hawking,Cargese}.
The latter
is defined as
\begin{equation}
\zeta(\nu) = {\rm Tr}\, {\hat H}^{-\nu} = {\rm Tr}\, e^{-\nu \ln 
{\hat H}}.
\label{zetanu}
\end{equation}
We can rewrite Eq.~(\ref{W1Hhat}) as
\begin{eqnarray}
W_q=-\,\frac{i}{2}\left[
\zeta'(0)+\ln (\mu^2)\,\zeta(0)\right],
\label{W1zeta}
\end{eqnarray}
where $\zeta'(\nu) = d\zeta(\nu)/d\nu$.
This expression can be regularized by analytic continuation
of $\zeta(\nu)$ in the parameter $\nu$ to make it 
and its first derivative well-defined at $\nu = 0$. The
dependence on the arbitrary constant $\mu$ can then be
absorbed into the definition of the renormalized
constants such as $G_N$ and $\Lambda$ that appear in the
Einstein action. As is well known, additional terms
quadratic in the Riemann tensor must be added to the Einstein
action to absorb all the dependence on $\mu$ 
through renormalization.

It is convenient to introduce an integral representation of
the generalized $\zeta$-function:
\begin{eqnarray}
\zeta(\nu)&=& {\rm Tr} {\hat H}^{-\nu}\nonumber\\
 & = & {\rm Tr}\left\{ \Gamma(\nu)^{-1}
\int_0^{\infty} i ds\,(i s)^{\nu-1}
e^{-i s ( {\hat H}-i\epsilon)}\right\}.
\label{zetaIntegral1}
\end{eqnarray}
The operator $H(x)$ of Eq.~(\ref{H}) may be regarded as
the Hamiltonian of a fictitious nonrelativistic particle
moving on a curved 4-dimensional hypersurface having
coordinates $x^{\mu}$. The operator $\exp(-i s {\hat H})$ is
the quantum mechanical evolution operator of this
fictitious particle, with $s$ being the ``proper time.''
This idea was introduced by Schwinger in the context
of quantum electrodynamics and applied by DeWitt in 
curved spacetime. The trace can be represented as
an integral over a complete set of position
eigenstates $| x \rangle$ of the fictitious particle:
\begin{equation}
\zeta(\nu)=\frac{1}{\Gamma(\nu)}\int d^4x \sqrt{-g}
\int_0^{\infty} i ds\,(i s)^{\nu-1}
\langle x \vert 
e^{-i s ( {\hat H}-i\epsilon)}\vert x \rangle .
\label{zetaK}
\end{equation}
The propagation amplitude for the particle to go from 
position $x'$ at time $s = 0$ to position $x$ at time $s$ is
\begin{equation}
K(x,x';is)\equiv \left<x\right.\vert
e^{-i s (\hat H-i\epsilon)}\vert \left. x'\right>.
\label{K}
\end{equation}
This propagator satisfies the Schr{\"o}dinger equation
\begin{equation}
i\frac{\partial}{\partial s}K(x, x';is)=H(x) K(x,x';is),
\label{SEq}
\end{equation}
and satisfies the ``initial'' condition
\begin{equation}
\lim_{s\to 0}K(x,x';is)=\delta(x,x').
\label{INITCOND}
\end{equation}
Note that what appears in Eq.~(\ref{zetaK}) is the coincidence limit
$x'\to x$ of Eq.~(\ref{K}). If one were to replace $is$ by $s$ in the
Schr{\"o}dinger equation, then it would become the equation governing
heat flow on the 4-dimensional hypersurface, in which case $K(x,x',s)$ 
is the heat kernel. Thus, the heat kernel and propagator are related by
analytic continuation in the proper time $s$.

The asymptotic expansion of the propagator in powers of $s$ can
be obtained from the Schr{\"o}dinger equation by iteration. The 
proper time $s$ has the dimensions of length squared, so the
coefficients of successive powers of $s$ in the series 
contain contracted products of increasing numbers of curvature 
tensors and/or covariant derivatives to balance the dimensions. 
The leading terms in this power series covariantly characterize 
the short wavelength behavior of the quantum field.
In four dimensions, the terms 
of order $s^{-2}$, $s^{-1}$, and $s^{0}$ 
in the series are
the ones that would give rise to ultraviolet (UV) divergences in 
the unregularized expression for the expectation value of 
the energy-momentum tensor. Those are the terms that are absorbed
through renormalization of the coupling constants of the 
curvature terms in the classical Einstein action (with 
counterterms quadratic in the Riemann curvature tensor included).

Writing the coincidence limit, $K(x,x;is)$, of the propagator 
(in four dimensions) in the form
\begin{eqnarray}
K(x,x;is)=\frac{i}{16 \pi^2 (i s)^2}\,e^{-i(m^2-i \epsilon) s}
             F(x,x;is),
\label{KF}
\end{eqnarray}
the proper-time series for $F(x,x;is)$ has the form
\begin{eqnarray}
F(x,x;is)=\sum_{j=0}^{\infty}(is)^jf_j,
\label{ASEXP}
\end{eqnarray}
where the first three terms are given by
\begin{eqnarray}
f_0&\equiv& f_0(x,x)=1,
\\
f_1&\equiv& f_1(x,x)=-\bar \xi R,
\label{f1}
\\
f_2&\equiv & f_2(x,x)=\frac{1}{2}\,{\bar \xi}{}^2 R^2+\frac{1}{180}\left[
\left(1-30 {\bar \xi}\right)\square R+
R_{\alpha \beta \gamma \delta}R^{\alpha \beta \gamma \delta}
-R_{\alpha \beta}R^{\alpha\beta}
\right],
\label{f2}
\end{eqnarray}
with $\bar \xi \equiv \xi -1/6$.
In general, this proper time series for $F(x,x;is)$ is
an asymptotic series that does not converge to the exact
solution for $F(x,x;is)$. Thus, it can only supply limited
information about the large $s$ behavior of $F$.

It turns out that the Schwinger-deWitt proper-time series 
given in Eq.~(\ref{ASEXP})
is sufficient to calculate one of the terms appearing in Eq.~(\ref{W1zeta}), 
namely the term involving $\zeta(0)$. In order to do so, we first have to
analytic continue Eq.~(\ref{zetaK}) to the value $\nu=0$. 
Substituting Eq.~(\ref{KF}) into Eq.~(\ref{zetaK}) and
performing three
integration by parts, one has:
\begin{eqnarray}
\zeta(\nu)&=&-\,\frac{i}{16 \pi^2(\nu-2)(\nu-1)\Gamma(\nu+1)}\int d^4x
\sqrt{-g}
\nonumber \\
& &\times
\int_0^\infty ids (is)^\nu
\frac{\partial^3}{\partial (is)^3}\left[
e^{-i (m^2- i\epsilon)s}F(x,x;is)
\right],
\label{zetaanalytic}
\end{eqnarray}
which is regular at $\nu=0$. In fact, one easily obtains:
\begin{eqnarray}
\zeta(0)&=&\frac{i}{32 \pi^2}\int d^4x\sqrt{-g}
\left.\left\{
\frac{\partial^2}{\partial (is)^2}\left[
e^{-i (m^2- i\epsilon)s}F(x,x;is)
\right]
\right\}\right|_{s=0}\nonumber \\
&=&\frac{i}{32 \pi^2}\int d^4x\sqrt{-g}\left(m^4-2 m^2 f_1+2 f_2\right).
\label{zeta0}
\end{eqnarray}
The remaining term in Eq.~(\ref{W1zeta}) involving
$\zeta'(0)$, on the other hand, cannot be 
calculated only based on the small-$s$ behavior of $F(x,x;is)$. Taking the
derivative of Eq.~(\ref{zetaanalytic}) with respect to $\nu$, and
applying the result to $\nu=0$, we have
\begin{eqnarray}
\zeta'(0)&=&\left(\frac{3}{2}+\gamma\right)\zeta(0)-
\frac{i}{32 \pi^2}\int d^4x\sqrt{-g}
\nonumber \\
& &\times
\int_0^{\infty}ids\, \ln(is)\frac{\partial^3}{\partial (is)^3}\left[
e^{-i(m^2-i\epsilon)s}F(x,x;is)
\right],
\label{zetaprime}
\end{eqnarray}
where $\gamma\equiv -\left.(d/dz)\ln\Gamma(z)\right|_{z=1}$ is
the Euler's constant.
As we shall show in Sec.~\ref{sec:general}, a most important 
contribution to $\zeta'(0)$ comes from the large-$s$
regime of the integrand of Eq.~(\ref{zetaprime}). 
Combining Eqs.~(\ref{zeta0}) and (\ref{zetaprime}) into Eq.~(\ref{W1zeta}),
we have the following expression for the one-loop effective action:
\begin{eqnarray}
W_q=-\,\frac{1}{64\pi^2}
\int d^4x \sqrt{-g}\left[
{\cal I}(0,+\infty)-\left(m^4-2 m^2 f_1+2 f_2\right)\ln{\tilde \mu}^2
\right],
\label{W1calI}
\end{eqnarray}
where, for given $0\leq \alpha <\beta \leq +\infty$,
\begin{eqnarray}
{\cal I}(\alpha,\beta)\equiv \int_\alpha^\beta ids \,
\ln(is)\frac{\partial^3}{\partial (is)^3}\left[
e^{-i(m^2-i\epsilon)s}F(x,x;is)
\right],
\label{calI}
\end{eqnarray}
and $\tilde \mu$ is such that $\ln {\tilde \mu}^2= 3/2+\gamma+ \ln \mu^2$.

Before analyzing in detail the properties of ${\cal I}(0,+\infty)$, we 
turn to some known results about the
coincidence limit of the propagator, $K(x,x;is)$. 
In the next section, we consider several cases in which 
the exact solution for $K(x,x;is)$ is known. Based on
these cases, we postulate a general form for the large-$s$
asymptotic form of $K(x,x;is)$. A similar asymptotic form 
is also suggested by the summation of the subset of terms in 
the proper-time series for $F(x,x;is)$ that involve one or more factors of
the scalar curvature $R$~\cite{PT,JP}.
This summation involves all powers of $s$, and thus may give
some information about the asymptotic form of $F(x,x;is)$ for
large $s$. In addition, the Feynman path integral solution
of the Schr{\"o}dinger equation (\ref{SEq}) to Gaussian 
order~\cite{BeP}
suggests a similar large-$s$
asymptotic behavior. It is the asymptotic form of $F(x,x;is)$
for large $s$ that determines the contributions of long wavelength 
infrared (IR) quantum fluctuations of the field $\phi$ to 
the functional integral for the effective action. If the
mass of the scalar particle is very small, it has a very large
Compton wavelength which may significantly influence the
long wavelength quantum fluctuations of the $\phi$ field. As we 
shall see, there are spacetimes in which this influence can cause
the expectation value of the energy-momentum tensor of
the free scalar field to become large.


\section{In search of the large-$S$ behavior of the propagator}
\label{sec:partcases}

It is well known that the series for $F(x,x;is)$ 
given in Eq.~(\ref{ASEXP}) is not convergent
in general. However, as conjectured by Parker and Toms~\cite{PT}, 
and later proved by Jack and Parker~\cite{JP},
Eq.~(\ref{ASEXP}) contains  a convergent sub-series 
involving powers of the scalar curvature $R$, with the property that
when it is summed, the (asymptotic) series that is left does {\it not} possess
any power of the scalar curvature (without derivatives applied to it). More
precisely, the series for $F(x,x;is)$ in powers of $s$ can 
be written (with ${\bar \xi} = \xi - 1/6$) as
\begin{eqnarray}
F(x,x;is) = e^{-i{\bar \xi}Rs}\sum_{j=0}^\infty
(is)^j{\bar f}_j,
\label{ASEXPRSUMMEDcoinc}
\end{eqnarray}
with the first three terms being
\begin{eqnarray}
{\bar f}_0&=&{\bar f}_0(x,x)=1,
\label{f0bar}
\\
{\bar f}_1&=&{\bar f}_1(x,x)=0,
\label{f1bar}
\\
{\bar f}_2&=&{\bar f}_2(x,x)=\frac{1}{180}\left[
\left(1-30 {\bar \xi}\right)\square R+
R_{\alpha \beta \gamma \delta}R^{\alpha \beta \gamma \delta}
-R_{\alpha \beta}R^{\alpha\beta}
\right],
\label{f2bar}
\end{eqnarray}
and the subsequent terms being free of any terms containing 
undifferentiated factors of $R$.
Obviously, fixing a particular background might allow one to write the 
contractions of the Riemann and Ricci tensors appearing in the ${\bar f}_j$
in terms of the scalar curvature $R$. For example, the symmetries of
de Sitter spacetime (a case that will be analyzed later)
make it possible to express any geometrical scalar quantity
in terms of $R$ alone. However, the dependence of ${\bar f}_j$ on $R$ 
would be different for different backgrounds. Similarly, there are
identities that depend on the dimension of spacetime that would allow
one to reexpress some of the coefficients ${\bar f}_j$ in such
a way that terms involving $R$ would appear~\cite{Xu,JP2}.
The expressions for the
${\bar f}_j$ that are valid for {\em general} metrics and in spacetimes
of {\em arbitrary} dimension do not include factors of $R$.
The form of the coincidence limit of the propagator given in 
Eq.~(\ref{ASEXPRSUMMEDcoinc}) is called
the $R$-summed, or the partially-summed, form of the propagator
or heat kernel.
The factor of $\exp(-i{\bar \xi}Rs)$ in Eq.~(\ref{ASEXPRSUMMEDcoinc})
sums a covariant and dimensionally-invariant set of terms to all
orders in $s$. Therefore, it may contain information about
the large-$s$ behavior of $K(x,x;is)$. Earlier, Bekenstein and
Parker~\cite{BeP}, using Fermi coordinates in curved spacetime, were
able to obtain the Gaussian approximation to the Feynman path 
integral solution of the Schr\"odinger equation (\ref{SEq}).
This approximation did not restrict $s$ to small values. The
result they obtained for $K(x,x';is)$ reduces in the coincidence
limit to the result obtained from the first term of the
series in Eq. (\ref{ASEXPRSUMMEDcoinc}). This gives further
support to the impression that the exponential factor
has relevance to the nonperturbative large-$s$ behavior 
of $K(x,x;is)$.

In the work of Parker and Raval~\cite{PR1,PR2345} 
they considered the effects of 
this nonperturbative term in $K(x,x;is)$ on the effective action. 
Here we clarify and extend their work in a number of respects.
First, by considering several exactly known heat kernels
we generalize the expression for the large-$s$ behavior 
of $K(x,x;is)$ that was suggested by the $R$-summed form and
by the Gaussian approximation. Second, we prove that the
growth in the vacuum expectation value of $T_{\mu\nu}$ that
occurs in certain spacetimes comes directly from the
large-$s$ asymptotic form of the propagator. Third, we
incorporate the possibility of dissipative processes into
our expression for the effective action. We also
summarize numerical solutions of the resulting semi-classical 
Einstein equations that we obtained in the FRW universe.
These numerical solutions include the contributions of higher 
derivatives of the Riemann tensor.

In order to arrive at a sufficiently general conjecture for
the large-$s$ asymptotic form of $K(x,x;is)$, 
let us analyze three particular cases where the propagator
$K(x,x';is)$ is known exactly, namely, de Sitter spacetime, the
Einstein static universe, and the linearly-expanding spatially-flat 
FRW universe (the latter one assuming conformal coupling $\xi=1/6$).
The exact Euclidean heat kernel $K_E(x,x';s)$
(which is the Euclideanized form
of the propagator $K$) in de Sitter spacetime for a massive 
scalar field
was calculated by Dowker and Critchley~\cite{DC1} and given by:
\begin{equation}
K_E(p,s) = \frac{1}{4\pi^2a^4}
\frac{d}{dp}
\sum_{j=0}^{+\infty}
(j+1/2)
\exp 
\left\{\frac{is}{a^2}
\left[9/4-(j+1/2)^2 \right]
\right\}
P_j(p) \; ,
\label{1}
\end{equation}
where $p\equiv\cos(\sqrt{2\sigma}/a)$, $\sigma = \sigma (x,x')$ is
half the square of the 
geodesic distance between $x$ and $x'$, $a$ is a constant related to
the scalar curvature by $R=12/a^2$, and
$P_j$ are the Legendre polynomials.
[Here we use signature $(-,+,+,+)$, so
that we have performed the substitution $\sigma \mapsto -\sigma$ in 
the results of
Ref.~\cite{DC1},
which uses signature $(+,-,-,-)$.]
Using $dP_j(p)/dp \vert_{p=1} = j(j+1)/2$, factoring out
the $j$-independent exponential factor, and multiplying Eq.~(\ref{1}) by
$i$ to go from the Euclidean to the Lorentzian metric, 
we obtain the exact de Sitter propagator in the coincidence limit:
\begin{eqnarray}
K(x,x\,;is) &=& e^{-i(m^2 + \xi R-i\epsilon)s}\; i K_E(1,s) \nonumber \\
&=& 
\frac{i}{8 \pi^2}\frac{R^2}{144}e^{-i(M^2-i\epsilon)s} \nonumber \\
& & \times
\sum_{j=0}^{\infty} j (j+1)(j+1/2)
\exp \left[ \frac{-iRs}{12}j(j+1) \right] \;,
\label{2}
\end{eqnarray}
where $M^2 \equiv m^2 + {\bar \xi}R$. (The exponential factor multiplying
$iK_E(1,s)$ sets the propagator to the notation 
used here.) Notice that the exponential for the term $j=0$ would be
exactly the same as the exponential obtained from the $R$-summed form of the 
propagator, $e^{-iM^2 s}$. However, this exponential in Eq.~(\ref{2}) appears
multiplying a factor proportional to $j$, and therefore gives no contribution
to the summation (possibly due to the high degree of symmetry of de Sitter 
spacetime). 
Since the effective action $W_q$ depends linearly
on the propagator $K$ [see Eqs.~(\ref{W1zeta}), (\ref{zetaK}), and 
(\ref{K})], we can investigate separately
the contributions to $W_q$ coming from each term
in the summation of Eq.~(\ref{2}):
\begin{eqnarray}
K_j(x,x\,;is)\equiv -\,
\frac{i}{16 \pi^2s^2}\,e^{-i(M_{(j)}^2-i\epsilon)s}\,{\cal R}_{2(j)}
(is)^2,\;
\;j\geq 0,
\label{deSitterKj}
\end{eqnarray}
with $M^2_{(j)}\equiv M^2+j(j+1)R/12$ and 
${\cal R}_{2(j)}\equiv j(j+1)(j+1/2)R^2/72$.
Note that 
\begin{eqnarray}
K(x,x\,;is)=\sum_{j=0}^\infty K_j(x,x\,;is).
\label{KsumKj}
\end{eqnarray}
As will become clear in the next section, the dominant term in
Eq.~(3.8) for large $s$ is the one with $j > 0$ for which
$\vert M_{(j)}^2\vert$ is the smallest. This value of $j$, which
we denote by $k$, will of course depend on the value of $\xi$. For
example, if $\xi > -1/6$, then $k = 1$ and 
$M_{(k)}^2 = m^2 + \xi R$.
More generally, we can say that (in the distributional sense)
the dominant behavior of the propagator in de Sitter spacetime,
for large $s$, is given by
\begin{eqnarray}
K(x,x\,;is) &\sim&-\,
\frac{i}{16 \pi^2s^2}\,e^{-i(M_{(k)}^2-i\epsilon)s}\,{\cal R}_{2(k)} (is)^2,
\label{deSitterlarges}
\end{eqnarray}
with $k \ge 1$.

Our second example, is the Einstein static universe.
In this case, the exact form of the coincidence limit
of the propagator is given by~\cite{DC2}
\begin{eqnarray}
K(x,x;is)&=&-\,\frac{i}{16 \pi^2 s^2}\,e^{-i(M^2-i\epsilon)s}
\left\{1+2
\sum_{j=1}^{\infty} e^{ij^2\pi^2 a^2/s} 
\left(
1+2ij^2\pi^2a^2/s
\right)
\right\},
\label{exactKEinstein}
\end{eqnarray}
where $a$ is related to the scalar curvature through $R=6/a^2$.
The summation in $j$ appearing in Eq.~(\ref{exactKEinstein})
is related to the fact that in the Einstein static universe there are
infinitely many geodesics connecting any point $x$ to itself.
The contribution to the propagator given by
the {\it direct path} connecting $x$ to itself (i.e., the trivial path),
is encompassed by the factor $1$ inside the curly brackets of 
Eq.~(\ref{exactKEinstein}). Note then that the Gaussian approximation for
the propagator gives the exact direct-path contribution in the
Einstein static universe.
If we restrict our attention to the
direct-path contribution, we have the large-$s$ regime of 
Eq.~(\ref{exactKEinstein}) given by
\begin{eqnarray}
K(x,x;is)&\sim&-\,\frac{i}{16 \pi^2 s^2}\,e^{-i(M^2-i\epsilon)s}.
\label{Einsteinlarges}
\end{eqnarray}
We also find that approximation of the infinite sum 
in Eq.~(\ref{exactKEinstein}) by an integral strongly 
suggests that the contributions of the indirect paths
sum to a quantity that grows much more slowly than $s^2$.
Therefore, we conclude that the rate of growth of the de Sitter 
heat kernel for large $s$ is faster than that of the heat 
kernel in the Einstein static universe.

As our last example, let us consider the
spatially-flat FRW universe that is expanding linearly 
in the FRW proper time coordinate. 
With conformal coupling, 
$\xi=1/6$, between  the field and the scalar curvature,
the propagator was calculated by Chitre and Hartle~\cite{CH} and
Charach and Parker~\cite{CP}. In the coincidence limit, it can 
be put into the form
\begin{eqnarray}
K(x,x;is)&=&
-\,\frac{i}{16 \pi^2 s^2}\,e^{-i(m^2-i\epsilon)s}
\left\{
1+\beta\,\frac{Rs}{6\pi}\int_{-\infty}^{+\infty}
dv \,\frac{e^{-6i(\cosh v)^2/(Rs)}}{(v+i\pi/2)^3}
\right\},
\label{exactKFRW}
\end{eqnarray}
where $\beta$ is an arbitrary constant.
Note again that the overall exponential factor appearing 
in Eq.~(\ref{exactKFRW}) is consistent with the 
$R$-summed form of the propagator, recalling that in the 
case of conformal coupling one has $\bar \xi =0$, i.e., $M^2=m^2$. 
Another point worth mentioning about Eq.~(\ref{exactKFRW})
is that for large values of $s$, the integral 
inside curly brackets gives no contribution to order
$s^j$ for $j \geq 0$. [This conclusion can be drawn by taking 
the limit $s\to \infty$ in the integrand in 
Eq.~(\ref{exactKFRW}) and noting that the integral vanishes
in this limit.] Therefore, we have that in the large-$s$ regime
the dominant behavior of Eq.~(\ref{exactKFRW})
can be written as
\begin{eqnarray}
K(x,x;is)&\sim&-\,\frac{i}{16 \pi^2 s^2}\,e^{-i(m^2-i\epsilon)s}\,
{\cal R}_\lambda (is)^\lambda,
\label{FRWlarges}
\end{eqnarray}
where $\lambda$ is some number smaller than $1$ and ${\cal R}_\lambda$
is a scalar quantity with the same dimension as $R^\lambda$.
For $\xi \ne 1/6$, we would expect $m^2$ in (\ref{FRWlarges})
to be replaced by $M^2$. 

Of the FRW metrics considered so far, the large $s$ asymptotic form of 
the heat kernel of de Sitter spacetime, Eq.~(\ref{deSitterlarges}), has 
the highest power of $s$ multiplying the exponentials that appear 
in all the examples. Therefore, in arriving at an {\it ansatz}
for the large $s$ asymptotic form of the heat kernel in a
general (but not pathological) FRW universe, we must include an 
exponential multiplied by a power of $s$ that is at least as large
as the power that appears in (\ref{deSitterlarges}).
Therefore, based on the results and discussion presented so far 
in this section, about the form of the
coincidence limit of the propagator, $K(x,x;is)$, 
in de Sitter spacetime, the Einstein static universe,
and the linearly-expanding spatially-flat FRW universe,
it seems reasonable to make the {\it ansatz} that, at least
in the four-dimensional FRW universes that will be considered in 
a later section, the general form of the
{\it dominant} term in the large-$s$ behavior of
$K(x,x;is)$ is given by 
Eq.~(\ref{KF}) with
\begin{eqnarray}
F(x,x;is)\sim {\cal R}_{n} \,e^{-i\chi_n R s}(is)^n
\label{ansatz}
\end{eqnarray}
for some integer $n$ (to be discussed in the next paragraph),
with $\chi_n$ being a dimensionless number and
${\cal R}_n={\cal R}_n(x)$ a scalar quantity
constructed from the metric $g_{\mu \nu}$
and having the same dimension as the $n$-th power of the 
scalar curvature $R$. It seems clear that the value of 
$\chi_n$ will depend on the constant $\xi$ that appears in
the field equation for $\phi$. It may also be related to the
topological properties of the spacetime because the 
large-$s$ asymptotic form of $F(x,x;is)$ could conceivably sense 
the large-scale structure of the spacetime. We will assume that
the value of $\xi$ is chosen such that $\chi_n$ is negative,
as negative values of $\chi_n$ are of interest for the cosmological
effect that we consider in a later section. The parameter that
determines the cosmological effect is the ratio 
$-m^2/ \chi_n\equiv \bar{m}^2$.

The particular cases analyzed here may give some hints on the values
of $n$ and ${\cal R}_n$. Since we are looking for the {\it general}
dominant large-$s$ behavior of $K(x,x;is$), 
the de Sitter case analyzed above seems to suggest that
$n$ is $2$. (It could be larger than 2, but our examples give no 
evidence of that.)
Then the Einstein static universe and the linearly expanding universe 
would be special cases in which subdominant asymptotic terms become 
dominant as a result of the vanishing of the coefficient of the 
dominant asymptotic term. 
Should this be true, then ${\cal R}_2$ must 
be a scalar that vanishes when calculated in the Einstein static 
universe and in the linearly-expanding spatially-flat FRW universe,
while being non-zero when calculated in de Sitter spacetime
(recalling also that its dimension is the same as the dimension 
of $R^2$).
It is not difficult to verify that
some of the candidates for the general 
expression of ${\cal R}_2$ in four dimensions are given by
the integrand of the {\it Gauss-Bonnet invariant}, 
\begin{eqnarray}
G\equiv 
R_{\alpha\beta\gamma\delta}R^{\alpha\beta\gamma\delta}-4R_{\alpha\beta}
R^{\alpha\beta}+R^2,
\label{GaussBonnet}
\end{eqnarray}
the second-order term in the $R$-summed form of
the Schwinger-deWitt proper-time series [see Eq.~(\ref{ASEXPRSUMMEDcoinc})],
\begin{eqnarray}
{\bar f}_2\equiv \frac{1}{180}\left[
(1-30 {\bar \xi})\square R+
R_{\alpha\beta\gamma\delta}R^{\alpha\beta\gamma\delta}-R_{\alpha\beta}
R^{\alpha\beta}
\right],
\label{R2f2bar}
\end{eqnarray}
and the scalar quantity
\begin{eqnarray}
{\cal S}\equiv
R_{\alpha\beta}R^{\alpha\beta}-\frac{R^2}{3}.
\label{3rdoption}
\end{eqnarray}
Actually, linear combinations of $G$, ${\bar f}_2$, and ${\cal S}$
are also candidates for the general expression of ${\cal R}_2$.

It is already clear from the work of Parker and Raval, that it is the 
exponential that appears in the asymptotic form of the propagator that 
is the key to causing an acceleration of the expansion of the universe.
In our numerical work, to be discussed in a later section, we will take
$n = 2$ and a particular choice for ${\cal R}_2$. We will also mention
numerical results we obtained by taking $n = 0$, which confirm that the
exponential is responsible for bringing about acceleration of the 
expansion. Although the case $n = 0$ was worth considering because
it involves fewer time derivatives of the metric, the case $n = 2$ 
appears to behave better, as will be discussed.

It is important to stress that the results we shall derive in the 
next section are {\it independent} of the specific form of the 
term ${\cal R}_n$ in Eq.~(\ref{ansatz})  
and of the assumption
that $n=2$ (as long as $n\geq 0$ 
and ${\cal R}_n$ is not identically zero for a general metric, 
which one immediately sees to 
be true from the cases we have considered). 
For this reason, in the next section 
we will use the generic {\it ansatz} for the
large-$s$ behavior of $F(x,x;is)$ given in
Eq.~(\ref{ansatz})
and only in later sections will we take $n=2$ and assume 
a particular form for ${\cal R}_2$.


\section{Large-{\it s}
asymptotic behavior of the propagator and 
nonperturbative infrared quantum effects}
\label{sec:general}

Returning to the form of the effective action given in Eq.~(\ref{W1calI}), 
we will split the quantity ${\cal I}(0,+\infty)$ in two terms
[see Eq.~(\ref{calI})]:
\begin{eqnarray}
{\cal I}(0,+\infty)=
{\cal I}_{reg}+{\cal I}_{IR},
\label{calIbreak}
\end{eqnarray}
where
\begin{eqnarray}
{\cal I}_{reg}&=&{\cal I}_{reg}({\lambda_{IR}})
\equiv {\cal I}(0,{\lambda_{IR}}),
\label{calIreg}
\\
{\cal I}_{IR}&=&{\cal I}_{IR}({\lambda_{IR}})\equiv {\cal I}
({\lambda_{IR}},+\infty),
\label{calIIR}
\end{eqnarray}
with ${\lambda_{IR}}$ some ``large'' but fixed
parameter with dimension $({\rm length})^2$.  
The contribution of long wavelength (IR) fluctuations of
the field $\phi$ to the effective action is given by the
integration over large $s$.

We will be most interested in the quantity ${\cal I}_{IR}$ and will
assume that ${\cal I}_{reg}$ is a well-behaved function of the
metric if $x$ is a non-singular point of the spacetime. This
assumption seems quite reasonable because we do not
expect the function 
$F(x,x;is)$ appearing in the definition of ${\cal I}_{reg}$ 
to give any problem in the {\it limited} interval of integration
$(0,\lambda_{IR})$. Moreover, from the Schwinger-deWitt
proper-time series we even know that $F(x,x;is)\to 1$ as
$s\to 0$. The situation is 
different, however, for the large-$s$  contribution ${\cal I}_{IR}$, 
as we will analyze next.

Combining Eqs.~(\ref{calIIR}), (\ref{calI}), and
(\ref{ansatz}), we have,
for sufficiently large ${\lambda_{IR}}$,
\begin{eqnarray}
{\cal I}_{IR}&\approx & {\cal R}_n\int_{\lambda_{IR}}^\infty
ids\,\ln(is)\frac{\partial^3}{\partial (is)^3}\left[
e^{-i(M_n^2-i\epsilon)s}\,(is)^n
\right]\nonumber \\
&=&(-1)^{n+1}{\cal R}_n \frac{\partial^n}{\partial (M_n^2)^n}\left[
\left(M_n^2-i\epsilon\right)^3
{\cal J}(\lambda_{IR},M_n^2)
\right],\;\;\;
\label{calIIRapprox}
\end{eqnarray}
where 
\begin{eqnarray}
M_n^2\equiv m^2 + \chi_n R
\label{M2N}
\end{eqnarray}
and the quantity ${\cal J}$ is defined by
\begin{eqnarray}
{\cal J}(\lambda_{IR},M_n^2)&\equiv& \int_{\lambda_{IR}}^\infty ids \,\ln(is)
\,e^{-i(M_n^2-i\epsilon)s}\nonumber \\
&=&i{\lambda_{IR}}
\int_1^\infty d\tilde s \left[\ln\tilde s+
\ln(i{\lambda_{IR}})\right]
 e^{-i{\lambda_{IR}}(M_n^2-i\epsilon)\tilde s}\nonumber \\
&=&\frac{\ln(i{\lambda_{IR}}) e^{-i{\lambda_{IR}}(M_n^2-i\epsilon)}}
{(M_n^2-i\epsilon)}
\nonumber \\
& &+i{\lambda_{IR}} \int_1^\infty
d\tilde s\,(\ln\tilde s)\, e^{-i{\lambda_{IR}}(M_n^2-i\epsilon)\tilde s},
\label{calJ}
\end{eqnarray}
with $\tilde s \equiv s/{\lambda_{IR}}$. Using Eq.~(4.358.1) of Ref.~\cite{GR},
the integral appearing in Eq.~(\ref{calJ}) can be evaluated:
\begin{eqnarray}
\int_1^\infty d\tilde s\,(\ln \tilde s)\,e^{-i{\lambda_{IR}}(M_n^2-i\epsilon)
\tilde s}
&=&\left.
\frac{\partial}{\partial \beta}\left\{\frac{\Gamma(\beta,i{\lambda_{IR}} 
M_n^2+\epsilon)}
{[i{\lambda_{IR}}
(M_n^2-i\epsilon)]^{\beta}}
\right\}\right|_{\beta=1}
\nonumber \\
&=&
-[i{\lambda_{IR}}(M_n^2-i\epsilon)]^{-1}\left\{
\gamma+\ln[i{\lambda_{IR}}(M_n^2-i\epsilon)]
\right\}
\nonumber \\
& &+\sum_{j=0}^{\infty}\frac{(-1)^j(i{\lambda_{IR}} M_n^2)^j}
{j!(j+1)^2},
\label{formula}
\end{eqnarray} 
where $\Gamma(\beta,\alpha)$ is the incomplete gamma function and
in passing from the first to the second line of Eq.~(\ref{formula}) we
have used the expansion of $\Gamma(\beta,\alpha)$ in powers of $\alpha$
(see Eq.~(8.354.2) of Ref.~\cite{GR}). Then,
Eq.~(\ref{calJ}) becomes
\begin{eqnarray}
{\cal J}(\lambda_{IR},M_n^2)&=&-(M_n^2-i\epsilon)^{-1}\left[
\gamma+\ln(M_n^2-i\epsilon)\right]\nonumber \\
& &+i{\lambda_{IR}} \sum_{j=0}^\infty
\frac{(-1)^j(i{\lambda_{IR}} M_n^2)^j}{(j+1)!}
\left[\frac{1}{j+1}-\ln(i{\lambda_{IR}})\right].
\label{calJfinal}
\end{eqnarray}
Finally, using Eq.~(\ref{calJfinal}) to evaluate Eq.~(\ref{calIIRapprox}),
we obtain for the dominant infrared contribution to the effective action
\begin{eqnarray}
{\cal I}_{IR}&\approx &(-1)^n{\cal R}_n
\frac{\partial^n}{\partial(M_n^2)^n}\left[
(M_n^2-i\epsilon)^2\ln(M_n^2-i\epsilon)
\right]\nonumber \\
& &+(-1)^{n}{\cal R}_n\left.\frac{\partial^n}{\partial (M_n^2)^n}
\right\{\gamma M_n^4
\nonumber \\
& &\left.
+(i{\lambda_{IR}})^{-2}\sum_{j=3}^{\infty}\frac{(-1)^j
(i{\lambda_{IR}} M_n^2)^j}{(j-2)!}\left[
\frac{1}{j-2}-\ln(i{\lambda_{IR}})
\right]
\right\}.
\label{calIIRfinalN}
\end{eqnarray}
Notice that the summation now starts at $j=3$.

There are a few points worth mentioning about Eq.~(\ref{calIIRfinalN}). First,
note that the series appearing above is (absolutely) convergent for any 
(finite) value of $({\lambda_{IR}} M_n^2)$. Moreover, the result of
any number of derivatives with 
respect to $M_n^2$ applied to this series is still (absolutely) convergent.
These facts imply that the 
first term on the right-hand-side of 
Eq.~(\ref{calIIRfinalN})
is the dominant one when
$M_n^2$ is sufficiently small. (In particular, for $n\geq 2$, the value of
this dominant term is unbounded when $M_n^2\to 0$, i.e., when $R\to -m^2/
\chi_n$). 
Notice that this dominant term in Eq.~(\ref{calIIRfinalN}) is {\it independent}
of ${\lambda_{IR}}$.
Thus, the effective action given in Eq.~(\ref{W1calI})
can be approximated by
\begin{eqnarray}
W_q&\approx &
W_q^{reg}+\frac{(-1)^{n+1}}{64 \pi ^2}\int d^4x\sqrt{-g}\,{\cal R}_n
\nonumber \\
& &\times
\frac{\partial^n}{\partial (M_n^2)^n}\left[
(M_n^2-i\epsilon)^2\ln\left(\frac{M_n^2-i\epsilon}{m^2}\right)\right],
\label{W1finalN}
\end{eqnarray}
where
\begin{eqnarray}
W_q^{reg}&\equiv&-\,\frac{1}{64 \pi^2}\left.\int d^4x\sqrt{-g}\right\{
{\cal I}_{reg}-\left(
m^4-2 m^2f_1+ 2 f_2
\right)\ln{\tilde \mu}^2\nonumber \\
& &
\left.
+(-1)^n{\cal R}_n\frac{\partial^n}{\partial (M_n^2)^n}
\right[
(\gamma +\ln m^2)M_n^4
\nonumber \\
& &\left.\left.
+(i{\lambda_{IR}})^{-2}\sum_{j=3}^\infty
\frac{(-1)^j(i{\lambda_{IR}} M_n^2)^j}{(j-2)!}\left(\frac{1}{j-2}-
\ln(i{\lambda_{IR}})\right)
\right]
\right\}
\label{W1reg}
\end{eqnarray}
denotes the part of the effective action whose value and
variation with respect to the metric $g_{\mu \nu}$ are well-behaved  
for any value of $M_n^2$. 

The positive infinitesimal $\epsilon$ in Eq.~(\ref{W1finalN}) was originally 
introduced to ensure convergence of the functional integral for $\exp(iW_q)$.
Dissipation through small interactions with external systems can be modeled by
allowing $\epsilon$ to be a finite quantity, small with respect to $m^2$. For
example, a long but finite lifetime of the particle associated with the
field $\phi$ may be modeled in this way.


\section{Renormalization and variation of the effective action}
\label{sec:renormvar}

Now, let us carry out the renormalization of the effective action 
given in 
Eq.~(\ref{W1finalN}). Notice that because we have used the $\zeta$-function
formalism, the effective action $W_q$ already at this stage
is free from  ``unphysical'' divergences.
However, $W_q$ given in Eq.~(\ref{W1finalN}) depends on the 
unobservable parameter $\tilde \mu$ through terms up to second
order in the curvature of the spacetime [see Eq.~(\ref{W1reg})].
Adding
a {\it bare} gravitational action containing terms up to 
second order in curvature,
\begin{eqnarray}
W_g\equiv \int 
d^4x \sqrt{-g}\left(
-2 \kappa \Lambda +\kappa R+\alpha_1 R^2+\alpha_2 R_{\mu\nu}R^{\mu \nu}
+\alpha_3 R_{\mu \nu \rho \sigma}R^{\mu \nu \rho \sigma}\right)
\label{bareWg}
\end{eqnarray}
with $\kappa$, $\Lambda$, $\alpha_1$, $\alpha_2$, $\alpha_3$ being bare
gravitational constants, we can absorb $\tilde \mu$ into the
definition of {\it observable} low-curvature gravitational constants
$\kappa_o$, $\Lambda_o$, $\alpha_{1o}$, $\alpha_{2o}$, $\alpha_{3o}$
with the result that the low-curvature limit of
the {\it total} action has the form
\begin{eqnarray}
W\equiv W_g+W_1
&\sim&
\int d^4x\sqrt{-g}\left(-2 \kappa_o 
\Lambda_o +\kappa_o R+\alpha_{1o} R^2\right.
\nonumber \\
& &\left. +\alpha_{2o} R_{\mu\nu}R^{\mu \nu}
+\alpha_{3o} R_{\mu \nu \rho \sigma}R^{\mu \nu \rho \sigma}+...
\right).
\label{totalWlowR}
\end{eqnarray}
Since this procedure was performed by Parker and Raval in 
Ref.~\cite{PR1}, using the Schwinger-deWitt
proper-time series, we will skip this detailed calculation here. 
What we will compute now is the form of the
total effective action after renormalization. In order to do so 
we note that the renormalization procedure described above has 
the net effect of replacing the bare gravitational constants 
in the bare gravitational action
by the observable ones (thus giving the renormalized
gravitational action), while adding to $W_q$ terms up to second order 
in the curvature in such a way to 
cancel the terms up to second order originally in 
the low-curvature expansion of $W_q$
(in this way giving the renormalized effective action). Therefore, 
we have:
\begin{eqnarray}
W=W_g+W_q=\left(W_g\right)_{ren}+\left(W_q\right)_{ren},
\label{totalWren}
\end{eqnarray}
with
\begin{eqnarray}
\left(
W_g
\right)_{ren}&=&
\int d^4x\sqrt{-g}\left( -2 \kappa_o 
\Lambda_o +\kappa_o R+\alpha_{1o} R^2\right.
\nonumber \\
& &\left. +\alpha_{2o} R_{\mu\nu}R^{\mu \nu}
+\alpha_{3o} R_{\mu \nu \rho \sigma}R^{\mu \nu \rho \sigma}\right)
\label{Wgren}
\end{eqnarray}
and assuming, for example, $n=2$ in Eq.~(\ref{W1finalN}) 
motivated by the discussion of 
Sec.~\ref{sec:partcases},
\begin{eqnarray}
\left(W_q\right)_{ren}&\approx & \left(
W_q^{reg}\right)_{ren}-\frac{1}{32 \pi^2}
\int d^4 x\sqrt{-g}\, {\cal R}_2 \ln\left(\frac{M_2^2-i\epsilon}{m^2}\right)
\nonumber \\
&\approx& \left(
W_q^{reg}\right)_{ren}
-\frac{i}{32\pi}\int d^4x\sqrt{-g}\,{\cal R}_2\,\Theta (-M_2^2)
\nonumber 
\\
& &-\frac{1}{64 \pi^2}
\int d^4 x\sqrt{-g}\, {\cal R}_2 \ln\left(\frac{M_2^4+\epsilon^2}{m^4}\right)
\label{W1ren}.
\end{eqnarray}
Recall that
$M_2^2=m^2+\chi_2 R$ [see Eq.~(\ref{M2N})] and that ${\cal R}_2$
is quadratic  in the curvature (see Sec.~\ref{sec:partcases}).
In passing from the first to the second line of Eq.~(\ref{W1ren}) 
we have used $\ln(M_2^2-i\epsilon)\approx\ln\vert M_2^2+
i\epsilon\vert +i\pi \Theta(-M_2^2)$, where
\begin{eqnarray}
\Theta(x)\equiv
\left\{
\begin{array}{ll}
0&,\;x<0,\\
1/2 &,\; x=0,\\
1&,\;x>0
\end{array}\right.
\label{Theta}
\end{eqnarray}
is the Heaviside step function.
The quantity $\left(W_q^{reg}\right)_{ren}$ in
Eq.~(\ref{W1ren}) is obtained by 
applying the renormalization procedure to $W_q^{reg}$ alone. 

The imaginary part of Eq.~(\ref{W1ren}) is related to 
particle production~\cite{PR2345}.
However, since the order of magnitude of the 
imaginary term written explicitly in Eq.~(\ref{W1ren})
could be comparable to the order of magnitude
of $\left(W_q^{reg}\right)_{ren}$, which may also
have an imaginary part that we are not obtaining explicitly, 
we will not analyze this phenomenon here.

Setting the observable gravitational constants $\alpha_{1o}$, 
$\alpha_{2o}$, and $\alpha_{3o}$ to 
zero (to reproduce the classical vacuum Einstein equations, 
with a cosmological constant, in the low-curvature
limit), the condition that the variation of the total effective action
vanishes for arbitrary variations of the metric gives us the
{\it semi-classical} vacuum Einstein equations:
\begin{eqnarray}
R_{\mu \nu}-\frac{1}{2}\,g_{\mu \nu}R
+\Lambda_og_{\mu \nu}
=8\pi G_N\left<T_{\mu \nu}\right>,
\label{sceeq}
\end{eqnarray}
where $G_N\equiv 1/(16 \pi \kappa_o)$ is Newton's constant, and 
the vacuum expectation value of the energy-momentum tensor of the
scalar field is defined through
\begin{eqnarray}
\left<
T^{\mu \nu}\right>\equiv
\frac{2}{\sqrt{-g}}\frac{\delta}{\delta g_{\mu \nu}}
\left(W_q\right)_{ren}.
\label{Tmunu}
\end{eqnarray}
As noted earlier, we will set $\Lambda_o = 0$ to see if
the expression for $\left< T^{\mu \nu}\right>$ could
explain the observed acceleration of the expansion of
the universe.

In order to numerically integrate the semi-classical
Einstein equations in an FRW universe, we must adopt 
an explicit form of $\left<T^{\mu \nu}\right>$.
For this purpose, we assume again $n=2$ and adopt an
expression for ${\cal R}_2$ that satisfies all the properties
discussed in Sec.~\ref{sec:partcases}. Namely, we let
\begin{eqnarray}
{\cal R}_2 =\alpha {\bar f}_2=\frac{\alpha}{180}
\left[
(1-30 {\bar \xi})\square R+
R_{\alpha\beta\gamma\delta}R^{\alpha\beta\gamma\delta}-R_{\alpha\beta}
R^{\alpha\beta}
\right],
\label{partR2}
\end{eqnarray}
where $\alpha$ is some dimensionless constant.
This is the term that appeared in the $R$-summed form of the 
heat kernel. It was used by Parker and Raval in their effective
action. Therefore, it allows us to directly compare our results
to the ones that were obtained from their analytic solution that
neglected derivatives of curvature invariants during the transition
to an expansion of the universe in which $R$ is constant. This
choice for ${\cal R}_2$ is sufficiently complicated to be representative
of other possibilities that contain higher derivatives of the metric.
In our numerical integration, we do not neglect higher derivatives
of the metric in the relevant terms, and we carefully analyze the
possibility of runaway solutions. As it would be too lengthy to
give the details here, we will present the full numerical analysis 
in a paper now in preparation~\cite{PVprep}; but here we give
our main conclusions and plot a representative numerical solution.

As we noted earlier, there are other possible choices of ${\cal R}_2$.
Our numerical analysis using the expression in Eq.~(\ref{partR2}) 
suggests that the transition from a classical expanding universe 
to an expansion with $R$ constant depends only on certain 
general features of ${\cal R}_2$, and thus may be a fairly robust 
generic feature of a wide class of asymptotic forms of the 
heat kernel. Therefore, it seems reasonable to also look at
effective actions suggested by other underlying theories that
have similar higher derivative terms. For example, string theory
leads to low-energy effective actions that contain higher derivative
terms analogous to those that appear in our low-energy 
effective action. Thus, our methods and analysis may be of interest 
to string theorists.

Combining Eqs.~(\ref{Tmunu}), (\ref{W1ren}), and (\ref{partR2}), and
using the variations presented in  Appendix~\ref{appa} [see
Eqs.~(\ref{varlog})-(\ref{varRicci2log})], we have
(after dropping the subscripts of $\chi_2$ and $M_2$ for simplicity):
\begin{eqnarray}
\left<
T^{\mu \nu}\right>&=&\left<T_{reg}^{\mu \nu}\right>
-\,\frac{\alpha}{16 \pi^2}
\left\{A_{(0)}^{\mu \nu}\ln\left(
\frac{M_\epsilon^4}{m^4}\right)+
\frac{\chi}{M_{\epsilon}^2} A_{(1)}^{\mu \nu}
+\frac{\chi^2}{M_{\epsilon}^4}A_{(2)}^{\mu \nu}
+\frac{\chi^3}{M_{\epsilon}^6}
A_{(3)}^{\mu \nu}
+\frac{\chi^4}{M_\epsilon^8}
A_{(4)}^{\mu \nu}
\right\},\nonumber \\
\label{evemtensor}
\end{eqnarray}
where we have defined 
\begin{eqnarray}
M_\epsilon^2 &\equiv& \sqrt{M^4+\epsilon^2}
=\sqrt{(m^2+\chi R)^2+\epsilon^2},
\label{Mepsilon}
\end{eqnarray}
and the regular tensors $A_{(j)}^{\mu \nu}$, $j=0,1,2,3,4$, are given
by
\begin{eqnarray}
A_{(0)}^{\mu \nu}&=&
\frac{1}{720}\left\{
2\nabla^\mu \nabla^\nu R-6 \square R^{\mu \nu}+
g^{\mu \nu}
\left(
\square R +R^{\alpha\beta\gamma\delta}
R_{\alpha\beta\gamma\delta}-R^{\alpha\beta}
R_{\alpha\beta}\right)\right.
\nonumber \\
& &\left.-
4R^{\mu\alpha\nu\beta}
R_{\alpha\beta}-4R^{\mu\alpha\beta\gamma}
R^\nu_{\;\,\alpha\beta\gamma}+8R^{\mu \alpha}R^\nu_{\;\,\alpha}
\right\},
\label{A(0)}
\end{eqnarray}
\begin{eqnarray}
A_{(1)}^{\mu \nu}&=&
\frac{1}{360}
\left(
\frac{M^2}{\sqrt{M^4+\epsilon^2}}
\right)
\left\{
(1+30{\bar \xi})g^{\mu \nu}\nabla^\alpha R\,\nabla_\alpha R
+12\nabla^\alpha R\,\nabla^{(\mu}R^{\nu)}_{\,\;\alpha}-12
\nabla^\alpha R\,\nabla_\alpha R^{\mu \nu}
\right.
\nonumber \\
& &\left.+2\left(
\nabla^\mu\nabla^\nu-g^{\mu \nu}\square -R^{\mu \nu}
\right)
\left[
2(1-30{\bar \xi})\square R+R^{\alpha\beta\gamma\delta}
R_{\alpha\beta\gamma\delta}-R^{\alpha\beta}R_{\alpha\beta}
\right]+2 R^{\mu \nu}\square R\right.\nonumber \\
& &\left.
-60 {\bar \xi}\nabla^\mu R\,\nabla^\nu R
+2g^{\mu \nu}R^{\alpha\beta}\nabla_\alpha\nabla_\beta R
-4R^{\alpha(\mu}\nabla^{\nu)}\nabla_\alpha R
-8R^{\mu \alpha\nu\beta}\nabla_\alpha\nabla_\beta R
\right\}
\label{A(1)}
\end{eqnarray}
\begin{eqnarray}
A_{(2)}^{\mu \nu}&=&
-\,\frac{1}{180}
\left(
\frac{M^4-\epsilon^2}{M^4+\epsilon^2}\right)
\left\{(1-30{\bar \xi})\left(
\nabla^\mu\nabla^\nu-g^{\mu \nu}\square -R^{\mu
\nu}\right)
\left(\nabla^\alpha R \,\nabla_\alpha R\right)
\right.
\nonumber \\
& &\left.+
\left[
\nabla^\mu\nabla^\nu R-g^{\mu \nu}\square R
+2\nabla^{(\mu}R\,\nabla^{\nu)}-2 g^{\mu \nu}\nabla^\lambda
R\,\nabla_\lambda
\right]\left[2(1-30 {\bar \xi})
\square R +R^{\alpha\beta\gamma\delta}
R_{\alpha\beta\gamma\delta}\right.\right.
\nonumber \\ 
& &
\left.\left.
-R^{\alpha\beta}
R_{\alpha\beta}
\right]+
\left(
R^{\mu \nu}g^{\alpha\beta}+R^{\alpha \beta}g^{\mu \nu}-2
R^{\alpha(\mu}
g^{\nu)\beta}-4 R^{\mu \alpha\nu\beta}
\right)\left(\nabla_\alpha R\,\nabla_\beta R\right)
\right\},
\label{A(2)}
\end{eqnarray}
\begin{eqnarray}
A_{(3)}^{\mu \nu}&=&
\frac{1}{90}
\left(
\frac{M^6-3\epsilon^2 M^2}{(M^4+\epsilon^2)^{3/2}}
\right)
\left\{
(1-30{\bar \xi})\left(
\nabla^\mu \nabla^\nu R-g^{\mu \nu}\square R\right)
\left(\nabla^\lambda R\,\nabla_\lambda R\right)
\right.
\nonumber \\
& &\left.
+
\left(\nabla^\mu R\,\nabla^\nu R-g^{\mu \nu}
\nabla^\lambda R\,\nabla_\lambda R\right)
\left[2(1-30 {\bar \xi})
\square R +R^{\alpha\beta\gamma\delta}
R_{\alpha\beta\gamma\delta}
-R^{\alpha\beta}
R_{\alpha\beta}
\right]\right.
\nonumber \\
& &\left.
+
2(1-30 {\bar \xi})
\nabla^\alpha R \,\nabla_\alpha
\left(\nabla^{\mu}R\,\nabla^{\nu}R-
g^{\mu \nu}\nabla^\lambda R\,\nabla_\lambda R
\right)
\right\},
\label{A(3)}
\end{eqnarray}
\begin{eqnarray}
A_{(4)}^{\mu \nu}&=&
-\,\frac{1}{30}
\left(
\frac{M^8-6 \epsilon^2 M^4 +\epsilon^4}{(M^4+\epsilon^2)^2}
\right)(1-30 {\bar \xi})\left(\nabla^\lambda R\,\nabla_\lambda R\right)
\left(\nabla^\mu R\, \nabla^\nu R-g^{\mu \nu}\nabla^\alpha
R\,\nabla_\alpha R\right).\nonumber \\
\label{A(4)}
\end{eqnarray}
The term $\left<T_{reg}^{\mu \nu}\right>$ in Eq.~(\ref{evemtensor})
stands for the part 
of the expectation value of the energy-momentum tensor coming from the
regular part of the effective action. 

Notice that the ratios  involving $M$ and $\epsilon$
appearing in parenthesis in each one of the
Eqs.~(\ref{A(1)})-(\ref{A(4)}) are
bounded functions of $M$. In fact, for $\epsilon$ very small, these
ratios are of order $1$ except for very particular values of $M$ for
which the ratios become close to zero. Notice also that in the limit 
$\epsilon \to 0$ all these ratios become equal to $1$.
In the following section we will apply the
expectation value given in Eq.~(\ref{evemtensor}) to an FRW spacetime
and analyze its cosmological consequences.


\section{Quantum scalar field in an FRW universe: the VCDM cosmological model}
\label{sec:vcdm}

Now we shall apply the results obtained in the previous sections to
a cosmological spacetime.
The goal is to show that the present accelerating
expansion of the universe may be explained as due to the 
nonperturbative-infrared form of the effective action calculated in 
Sec.~\ref{sec:renormvar}. Such a model for the ``dark energy'' was first
proposed by Parker and Raval and it is known as the VCDM or 
vacuum metamorphosis
model~\cite{PR1}-\cite{PKV}.

As an idealization, we will consider our universe
as being described by a spatially-flat FRW spacetime, with line element
\begin{eqnarray}
ds^2=-dt^2+a(t)^2\left(
dx_1^2+dx_2^2+dx_3^2
\right),
\label{ds2FRW}
\end{eqnarray}
filled with 
non-interacting matter, radiation, and a scalar field with
zero expectation value, $\left< \phi \right>=0$, and
a expectation value for its energy-momentum tensor 
$\left<
T^{\mu \nu}\right>$
given by Eqs.~(\ref{evemtensor})-(\ref{A(4)}). The 
semi-classical Einstein equations (with zero cosmological constant) 
then read
\begin{equation}
R^{\mu \nu}-\frac{1}{2}\,g^{\mu \nu}R=
8 \pi G_N \left(T_r^{\mu \nu}+T_m^{\mu \nu}+\left<T^{\mu \nu}\right>
\right)\;,
\label{scee}
\end{equation}
where $T_r^{\mu \nu}=\rho_r \left( g^{\mu \nu}+4 u^\mu u^\nu
\right)/3$ and $T_m^{\mu \nu}=\rho_m u^\mu u^\nu $, with
$u^\mu$ being the future-pointing,
normalized vector field orthogonal to the 
homogeneous and isotropic hypersurfaces, and $\rho_r$ and $\rho_m$
being the radiation and matter energy densities, respectively,
as measured by the family of geodesic
observers with four-velocity $u^\mu$
(comoving observers). Moreover,  $\rho_r$ and $\rho_m$ are constant
over each homogeneous and isotropic hypersurface.

The symmetries of the FRW spacetime greatly simplify the task of 
solving Eq.~(\ref{scee}). It is not difficult to see that 
the most general form of a rank-two tensor field
${\cal T}^{\mu \nu}$ that is consistent with the FRW symmetries is
\begin{equation}
{\cal T}^{\mu \nu}= {\cal A} u^\mu u^\nu +{\cal B} \left(
g^{\mu \nu}+u^\mu u^\nu \right)\;,
\label{gf}
\end{equation}
with ${\cal A}\equiv {\cal T}^{\mu \nu}u_\mu u_\nu$ 
and ${\cal B}\equiv \left(
{\cal T}^\mu_\mu+{\cal A}\right)/3$ being scalar functions that are 
constant over each homogeneous and isotropic hypersurface,
i.e., 
\begin{equation}
\nabla_\mu {\cal A}=-u_\mu u^\nu \nabla_\nu {\cal A}\;\;\; 
\text{and}\;\;\;
\nabla_\mu {\cal B}=-u_\mu u^\nu \nabla_\nu {\cal B}\;.
\label{spatialconst}
\end{equation}
Take, then,
${\cal T}^{\mu \nu}$ to be the difference between the 
left-hand-side (l.h.s.)
and the right-hand-side (r.h.s.) of Eq.~(\ref{scee}), which 
in an FRW spacetime clearly exhibits the form presented in 
Eq.~(\ref{gf}). This shows that the usually ten (semi-classical) 
Einstein equations, ${\cal T}^{\mu \nu}=0$, 
are reduced to two equations, 
${\cal A}={\cal B}=0$, in an FRW spacetime. Moreover,
considering
that $\nabla_\mu {\cal T}^{\mu \nu}=0$ is also satisfied [for
both sides of Eq.~(\ref{scee}) satisfy this condition], we have,
after using Eq.~(\ref{spatialconst}) and the fact that $u^{\mu}$ is
a geodesic field,
\begin{eqnarray}
0=\nabla_\mu {\cal T}^{\mu \nu}=u^{\nu}\left[
\nabla_\mu \left({\cal A} u^\mu\right)
+{\cal B} \nabla_\mu u^\mu\right]\;,
\label{bi}
\end{eqnarray}
which tells us that equation  
$\nabla_\mu {\cal T}^{\mu \nu}=0$ is equivalent to
${\cal B}=\left(\nabla_\alpha u^\alpha \right)^{-1}
\nabla_\mu \left({\cal A} u^\mu\right)$, {\it provided} 
$\nabla_\alpha u^\alpha \neq 0$. 
In this case, ${\cal A}= 0$ and 
$\nabla_\mu {\cal T}^{\mu \nu}=0$ 
imply that ${\cal B}=0.$
Then, summarizing what we have
shown so far, solving
Eq.~(\ref{scee}) in an FRW spacetime is equivalent to solving
${\cal A}={\cal B}=0$, which, in turn, is equivalent to
solving ${\cal A}=0$ {\it and} $\nabla_\mu {\cal T}^{\mu \nu}=0$
(assuming $\nabla_\alpha u^\alpha \neq 0$). In other words,
we are left with the problem of solving
\begin{eqnarray}
R^{\mu \nu}u_\mu u_\nu+\frac{1}{2}\,R
=
8 \pi G_N \left(\rho_r+\rho_m+\left<T^{\mu \nu}\right>u_\mu u_\nu
\right)\;,
\label{eq1}
\end{eqnarray}
\begin{eqnarray}
\nabla_\mu \left( T_r^{\mu \nu}+T_m^{\mu \nu}\right)=0\;.
\label{eq2}
\end{eqnarray}
[To obtain Eq.~(\ref{eq2}) we have used the fact that
$\nabla_\mu (R^{\mu \nu}-g^{\mu \nu}R/2)=
\nabla_\mu \left< T^{\mu \nu} \right>=0$ 
are identities since both $\sqrt{-g}(R^{\mu \nu}-g^{\mu \nu}R/2)$ and  
$\sqrt{-g}\left< T^{\mu \nu} \right>$ 
are given  by the functional derivative
of an action with respect to the metric.] Eq.~(\ref{eq2}) is simple 
to solve analytically, but the Eq.~(\ref{eq1}) can only be 
solved numerically. 

\begin{figure}
\epsfig{file=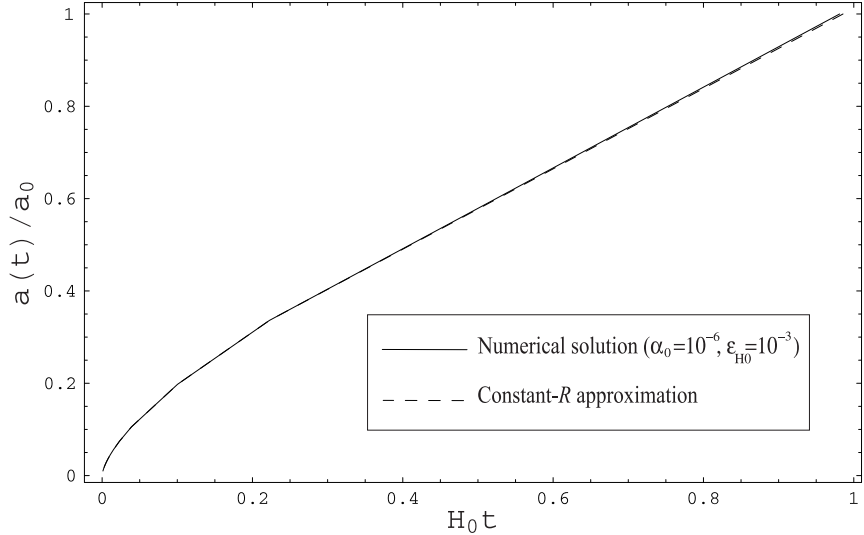,angle=0,width=0.7\linewidth}
\caption{Plot of the numerical solution 
(solid line) and constant-$R$ approximation (dashed line)
for the scale parameter $a(t)$, as a function of time $t$,
of a universe with matter and radiation content
given by $\Omega_{m0}=0.34$ and $\Omega_{r0}=8.33 \times 10^{-5}$. 
In order to 
facilitate the numerical analysis, we have used 
$\epsilon_{H0}\equiv \epsilon /H_0^2=10^{-3}$ and 
$\alpha_0 \equiv \alpha G_N H_0^2 =10^{-6}$.  Also, 
we used ${\bar m}/H_0=3.26$.}
\label{fig:axt}
\end{figure}

Space does not permit us to go into the details~\cite{PVprep}
of solving
numerically the ordinary differential equation for $a(t)$ 
obtained from
Eq.~(\ref{eq1}). Here, we show representative plots, and
summarize the main conclusions of the numerical calculations.
In Fig.~\ref{fig:axt} we show the result for the scale 
parameter $a(t)$ for a universe with present value of 
matter and radiation energy densities such that
$\Omega_{m0}\equiv \rho_m(t_0)/\rho_c=0.34$ and $\Omega_{r0}\equiv 
\rho_{r}(t_0)/\rho_c=8.33\times 10^{-5}$, with $\rho_c\equiv 
3H_0^2/(8\pi G_N)$,
$H_0$ being the present value of the Hubble constant. 
We compare this result (solid line) with the approximation for 
$a(t)$ given by the earlier version of the VCDM model, where 
a constant-scalar curvature stage follows the usual matter 
dominated phase of the universe (dashed line). 
We can see that the ``constant-$R$ approximation'' in fact provides 
a very good (analytical) approximation to the numerical 
solution $a(t)$. The mass of the scalar
field was chosen to be such that ${\bar m}/H_0=3.26$, and 
in order to facilitate the numerical analysis a non-zero value for 
$\epsilon$ was assumed (see discussion at the end of Sec.~\ref{sec:general} for
the physical meaning of $\epsilon$). Also for numerical reasons, 
the value of the dimensionless parameter 
$\alpha_0\equiv \alpha G_N H_0^2$, on which 
$a(t)$ depends through the quantum energy-momentum tensor
$\left< T^{\mu \nu} \right>$, was taken to be much larger than 
its physical value $\alpha_0\approx 10^{-122}$. 
This, however, should not be a problem since we
find that the smaller the
value of $\alpha_0>0$, the closer the numerical solution
for $a(t)$ becomes to the constant-$R$ 
approximation.

\begin{figure}
\epsfig{file=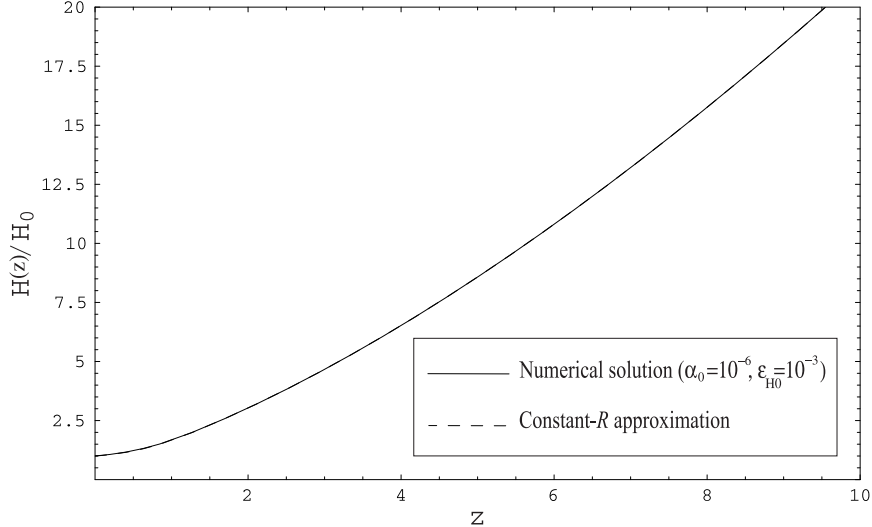,angle=0,width=0.7\linewidth}
\caption{Plot of the numerical solution 
(solid line) and constant-$R$ approximation (dashed line)
for the Hubble parameter $H(z)$, as a function of redshift $z$,
of a universe with matter and radiation content
given by $\Omega_{m0}=0.34$ and $\Omega_{r0}=8.33 \times 10^{-5}$. In order to 
facilitate the numerical analysis, we have used 
$\epsilon_{H0}\equiv \epsilon /H_0^2=10^{-3}$ and 
$\alpha_0 \equiv \alpha G_N H_0^2 =10^{-6}$.  Also, we used 
${\bar m}/H_0=3.26$.}
\label{fig:Hxz}
\end{figure}

In Fig.~\ref{fig:Hxz} we plot the Hubble parameter $H(z)$ 
(normalized
by its present value $H_0$) as a function of red-shift $z$. 
Note again that 
the numerical solution for $H(z)$ (solid line) 
is very well approximated by the constant-$R$
approximation (dashed line).

\begin{figure}
\epsfig{file=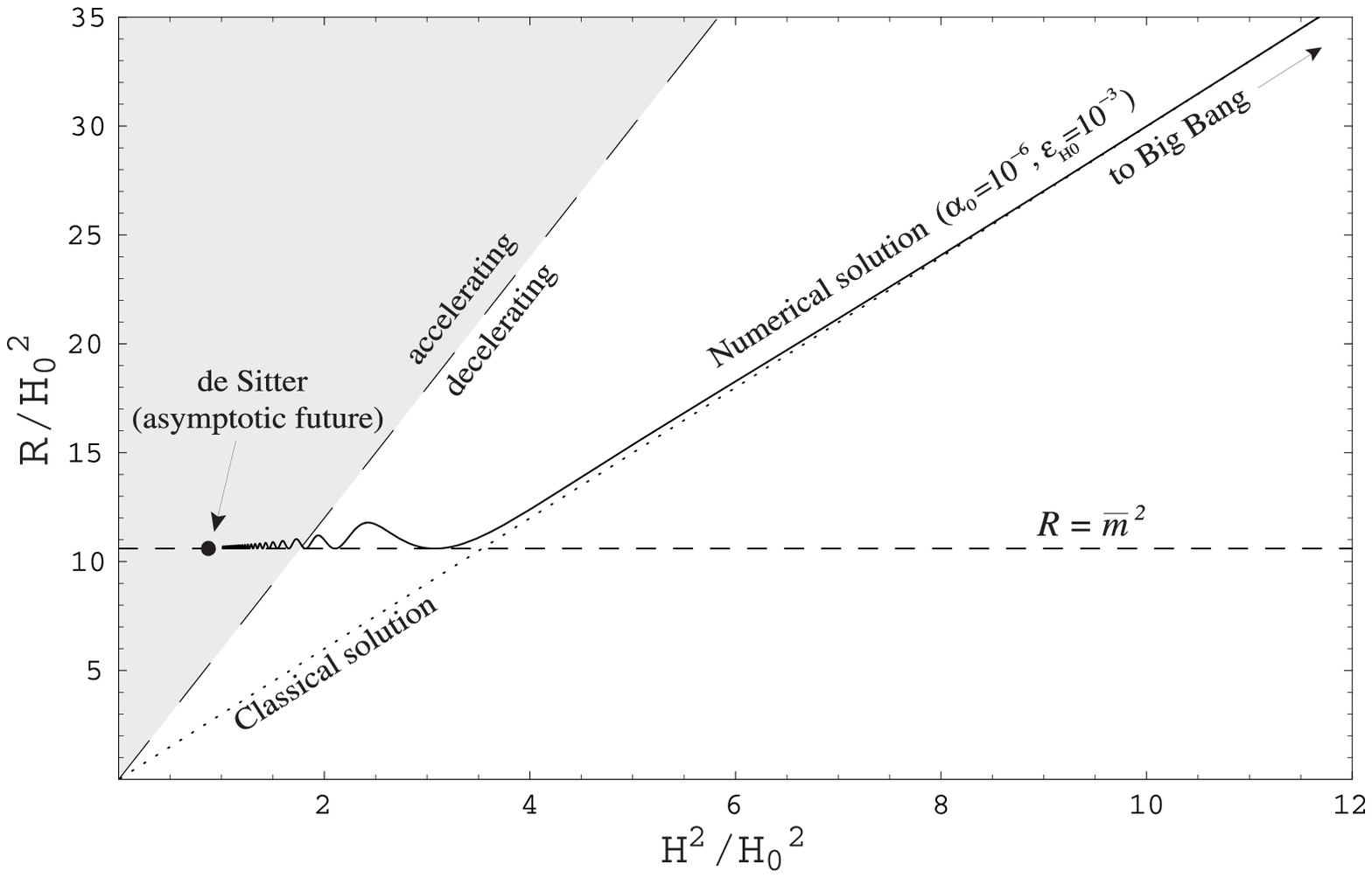,angle=0,width=0.7\linewidth}
\caption{Plot of the evolution of the square of 
the Hubble parameter $H^2$ and
scalar curvature $R$ of a universe  with matter 
and radiation content given by $\Omega_{m0}=0.34$ 
and $\Omega_{r0}=8.33 \times 10^{-5}$. Note that 
after a classical period of expansion (represented 
by the dotted line), the universe enters an era dominated 
by the energy-momentum tensor of the quantum
scalar field that prevents the scalar curvature $R$ 
from dropping below the value ${\bar m}^2$. 
Eventually, the universe enters a stage of accelerating 
expansion that leads to an asymptotic exponentially-fast 
expansion. In order to facilitate the numerical analysis, 
we have used 
$\epsilon_{H0}\equiv \epsilon /H_0^2=10^{-3}$ and 
$\alpha_0 \equiv \alpha G_N H_0^2 =10^{-6}$. 
Also, we used ${\bar m}/H_0=3.26$.}
\label{fig:phasediagram}
\end{figure}

Finally, in Fig.~\ref{fig:phasediagram} we present 
the numerical evolution of the universe (solid line)
in a diagram  showing its scalar curvature $R$ and 
the square of the Hubble parameter $H^2$. Notice that 
after spending some time in the 
matter-dominated stage (represented by the
dotted line), the universe makes a transition (not as sharp as 
in the constant-$R$ approximation) to an era dominated by the 
expectation value of the energy-momentum tensor 
of the quantized scalar field. During this latter era, 
the universe enters a period of accelerating 
expansion that lasts forever (in the spatially-flat FRW 
case) and that approaches, asymptotically in the future,
an exponentially fast expansion with Hubble parameter
$H\to{\bar m}/(2\sqrt{3})$.

It is important to mention that in numerically solving 
the semi-classical Einstein equations presented here,
we have taken into account the higher derivative terms.
Note that Eq.~(\ref{eq1}) leads
to a fifth-order differential equation for $a(t)$. For this reason, 
it is natural to expect that there exist
solutions to Eq.~(\ref{eq1}) that are not physically acceptable 
(e.g., runaway solutions). Notwithstanding, we find that
taking initial conditions such that $a(t)$ behaves
classically (e.g., describing a matter-dominated universe)
during some period of time in the past is enough to select only 
physically acceptable solutions, {\it all} of them evolving, 
eventually, to a phase of {\it accelerating expansion}. Since the
vacuum energy density and pressure are negligible 
at early times, the classical initial conditions are the natural 
ones to impose.
Initial conditions sufficiently far away from the classical ones
in the past do appear to give rise to unphysical solutions, but it
is remarkable that the class of natural initial conditions 
evidently gives only physically reasonable solutions.

As noted earlier, we also numerically integrated the
semiclassical Einstein equations that result from using
the {\it ansatz} of Eq.~(\ref{ansatz}) with $n = 0$. These equations
are simpler to solve numerically because they contain fewer time
derivatives of the metric. We found that, as expected, the
rate of change of the scalar curvature $R$ decreases as
$R$ approaches the value $\bar{m}^2$ from above. However,
instead of making a transition to an expansion in which
$R$ remains close to $\bar{m}^2$, the scalar curvature
effectively bounces off the value $\bar{m}^2$ and evolves 
toward increasing values of $R$. The figures that we show
are all for the case of $n = 2$, which seems to be of more
physical interest.

\section{Discussion}
\label{sec:discussion}

We have reconsidered the theory of the vacuum metamorphosis
transition that occurs in the vacuum cold dark matter
cosmological model from a manifestly nonperturbative
point of view.
We showed first that the terms in the vacuum energy-momentum tensor
that become large when $R$ is close to the value 
$\bar{m}^2= -m^2/\chi$ derive
from the large-$s$ asymptotic behavior of the heat kernel. Then by 
examining the large-$s$ asymptotic form of the exact heat kernel 
in the de Sitter, Einstein static, and linearly expanding FRW 
universes, we arrived at
a reasonable {\it ansatz} for the dominant asymptotic behavior of 
the heat kernel in a general FRW universe. The key feature of the 
heat kernel that ultimately causes the vacuum energy density 
and pressure to become large is the presence of a term 
of the form $\exp(-i\chi R s)$ in the 
large-$s$ regime of 
the heat kernel. Our approach is manifestly non-perturbative 
in $s$ because it involves the large-$s$ asymptotic 
form of the heat kernel, and does not involve summation of a series
in powers of $s$. [There is also a factor $\exp(-i m^2 s)$ in the
heat kernel
that is exact, and is thus valid for any nonzero 
mass $m$ and for all values of $s$.]

We then obtained the explicit renormalized
expression for the terms in the vacuum expectation value of 
$T_{\mu\nu}$ that become large when the scalar curvature $R$ 
is close to $\bar{m}^2$. 
These terms involve up to fifth-order time derivatives of the
FRW scale factor $a(t)$.
We also introduced a small parameter 
$\epsilon$ (of dimension $m^2$) that phenomenologically describes 
possible weakly dissipative effects, and softens the growth of
$T_{\mu\nu}$ as $R$ approaches $\bar{m}^2$. 

Finally, we adopted
a specific form for the invariant factor, quadratic in the
Riemann tensor, that multiplies the exponentials in the heat kernel,
and we numerically integrated the Einstein equations in a
spatially flat FRW universe, with the source consisting of 
the energy-momentum tensor of classical matter 
and radiation and the expectation value 
of $T_{\mu\nu}$ of the quantized scalar field $\phi$. 
We numerically evolved the terms having
the higher time derivatives of $a(t)$. We found that
if the universe in the past evolved classically (when the vacuum 
expectation values of $T_{\mu\nu}$ are negligible), then there
are no runaway or unphysical solutions. All solutions with
such classical initial conditions undergo a transition to 
an accelerating expansion that approaches the de Sitter
expansion at late times.

We also noted (see the Introduction) that the theory
considered here may provide a mechanism for early
inflation, which could be caused by heavy bosons, or by
the very same ultralight acceletron that may be
responsible for the presently observed acceleration
of the universe.

\begin{flushleft}
{\bf{\large Acknowledgments}}
\end{flushleft}

This work was supported by NSF grant PHY-0071044. 
D.~V.~thanks Atsushi Higuchi for very helpful discussions 
during the early stages of this work.
L.~P.~thanks Daniel Chung and Atsushi
Higuchi for helpful discussions of graviton self-energy
contributions to the mass of the scalar field.

\appendix
\setcounter{equation}{0}
\section{\label{appa} Variations}

We present here the list of the variations that were used in 
Sec.~\ref{sec:renormvar} (recall that $M^2 \equiv m^2 +\chi R$
and $M_\epsilon^2\equiv \sqrt{M^4+\epsilon^2}$):
\begin{eqnarray}
\delta
\left(
\int d^nx
\sqrt{\left|g\right|}\;
\ln\left(\frac{M_\epsilon^4}{m^4}\right)
\right)
=\int d^nx
\sqrt{\left|g\right|}\left\{
\frac{1}{2}\,g^{\mu \nu}\ln\left(
\frac{M_\epsilon^4}{m^4}
\right)\right.
\nonumber \\
\left.+
2\chi \left(\nabla^\mu\nabla^\nu-g^{\mu \nu}\square -R^{\mu \nu}\right)
\left(\frac{M^2}{M_\epsilon^4}\right)
\right\}\delta g_{\mu \nu}\;,
\label{varlog}
\end{eqnarray}
\begin{eqnarray}
\delta
\left(
\int d^nx
\sqrt{\left|g\right|}\;R\,
\ln\left(\frac{M_\epsilon^4}{m^4}\right)
\right)
=\int d^nx
\sqrt{\left|g\right|}\left\{
\frac{R}{2}\,g^{\mu \nu}\ln\left(\frac{M_\epsilon^4}{m^4}\right)
\right.
\nonumber \\
\left.
+\left(\nabla^\mu\nabla^\nu-g^{\mu \nu}\square -R^{\mu \nu}\right)
\left[
\ln\left(\frac{M_\epsilon^4}{m^4}\right)+\frac{2\chi R M^2}
{M_\epsilon^4}
\right]
\right\}\delta g_{\mu \nu}\;,\nonumber \\
\label{varRlog}
\end{eqnarray}
\begin{eqnarray}
\delta
\left(
\int d^nx
\sqrt{\left|g\right|}\;
R^2 \ln\left(\frac{M_\epsilon^4}{m^4}\right)
\right)
=\int d^nx
\sqrt{\left|g\right|}\left\{
\frac{R^2}{2}\,g^{\mu \nu}\ln\left(\frac{M_\epsilon^4}{m^4}\right)
\right.
\nonumber \\
\left.
+2\left(\nabla^\mu\nabla^\nu-g^{\mu \nu}\square -R^{\mu \nu}\right)
\left[
R\ln\left(\frac{M_\epsilon^4}{m^4}\right)+\frac{\chi R^2 M^2}
{M_\epsilon^4}
\right]
\right\}\delta g_{\mu \nu}\;,\nonumber \\
\label{varR2log}
\end{eqnarray}
\begin{eqnarray}
\delta
\left(
\int d^nx
\sqrt{\left|g\right|}\;
\ln\left(\frac{M_\epsilon^4}{m^4}\right)\square R
\right)
=\int d^nx
\sqrt{\left|g\right|}\left\{
\frac{\chi M^2}{M_\epsilon^4}\,g^{\mu \nu}\nabla^\alpha R\,
\nabla_\alpha R\right.\nonumber \\
\left.+
\left(\nabla^\mu\nabla^\nu-g^{\mu \nu}\square -R^{\mu \nu}\right)
\left[
\frac{2\chi M^2}{M_\epsilon^4}\square R +
\square
\ln\left(
\frac{M_\epsilon^4}{m^4}
\right)
\right]
\right\}\delta g_{\mu \nu}\;,
\label{varboxlog}
\end{eqnarray}
\begin{eqnarray}
\delta
\left(
\int d^nx
\sqrt{\left|g\right|}\;
R^{\mu \nu \rho \sigma} R_{\mu \nu \rho \sigma}
\ln\left(\frac{M_\epsilon^4}{m^4}\right)
\right)
=\int d^nx
\sqrt{\left|g\right|}\left\{
\frac{1}{2}\,g^{\mu \nu}
R^{\alpha\beta\gamma\delta}R_{\alpha\beta\gamma\delta}
\ln\left(\frac{M_\epsilon^4}{m^4}\right)
\right.\nonumber \\
\left.
2\chi \left(\nabla^\mu\nabla^\nu-g^{\mu \nu}\square -R^{\mu \nu}\right)
\left(
\frac{M^2}{M_\epsilon^4}
R^{\alpha\beta\gamma\delta}R_{\alpha\beta\gamma\delta}
\right)-2
R^{\mu\alpha\beta\gamma}R^\nu_{\;\,\alpha\beta\gamma}
\ln\left(\frac{M_\epsilon^4}{m^4}\right)
\right.\nonumber
\\
\left.-4\nabla_{(\alpha}\nabla_{\beta)}
\left[R^{\mu\alpha\nu\beta}\ln\left(\frac{M_\epsilon^4}
{m^4}\right)
\right]
\right\}
\delta g_{\mu \nu}\;,
\label{varRiemann2log}
\end{eqnarray}
\begin{eqnarray}
\delta
\left(
\int d^nx
\sqrt{\left|g\right|}\;
R^{\mu \nu} R_{\mu \nu}
\ln\left(\frac{M_\epsilon^4}{m^4}\right)
\right)
=\int d^nx
\sqrt{\left|g\right|}\left\{
\frac{1}{2}\,g^{\mu \nu} R^{\alpha \beta}R_{\alpha\beta}
\ln\left(\frac{M_\epsilon^4}{m^4}\right)
\right.
\nonumber \\
\left.+2\chi
\left(\nabla^\mu\nabla^\nu-g^{\mu \nu}\square -R^{\mu \nu}\right)
\left(
\frac{M^2}{M_\epsilon^4}
R^{\alpha\beta}R_{\alpha\beta}
\right)-2R^{\mu\alpha}R^\nu_{\;\,\alpha}
\ln\left(\frac{M_\epsilon^4}{m^4}\right)
\right.\nonumber \\
\left.
+2\nabla_\alpha \nabla^{(\mu}\left[
R^{\nu)\alpha}
\ln\left(\frac{M_\epsilon^4}{m^4}\right)
\right]
-\square
\left[
R^{\mu\nu}
\ln\left(\frac{M_\epsilon^4}{m^4}\right)
\right]-g^{\mu \nu}
\nabla_\alpha
\nabla_\beta
\left[
R^{\alpha\beta}
\ln\left(\frac{M_\epsilon^4}{m^4}\right)
\right]
\right\}
\delta g_{\mu \nu}.\nonumber \\
\label{varRicci2log}
\end{eqnarray}
Notice that one can obtain the variation of the factors multiplying
the logarithmic term in the l.h.s. of the
Eqs.~(\ref{varlog})-(\ref{varRicci2log})
as the particular case where $\chi=0$.


\end{document}